\documentclass[a4paper,preprintnumbers,floatfix,superscriptaddress,pra,twocolumn,showpacs,notitlepage,longbibliography]{revtex4-1}

\usepackage[utf8]{inputenc}
\usepackage[T1]{fontenc}
\usepackage[sc,osf]{mathpazo}
\usepackage{amsmath, amsthm, amssymb,amsfonts,mathbbol,amstext}
\usepackage{graphicx}
\usepackage{dcolumn}
\usepackage{bm}
\usepackage{bbm}
\usepackage{hyperref}
\usepackage{mathtools}
\usepackage{comment}
\usepackage{color}
\usepackage{multirow}
\usepackage{pgf,tikz}
\usepackage{mathrsfs}
\usetikzlibrary{arrows}
\usepackage{soul,xcolor}

\definecolor{Blue}{rgb}{0,0,1}
\definecolor{Red}{rgb}{1,0,0}
\definecolor{Green}{rgb}{0,1,0}
\definecolor{darkgreen}{rgb}{0,.7,0}
\definecolor{Purp}{rgb}{.2,0,.2}
\definecolor{white}{rgb}{1,1,1}

\def\1{\mathbf{1}}
\def\0{\mathbf{0}}

\DeclareMathOperator{\Tr}{Tr}

\newcommand{\ket}[1]{| #1 \rangle}
\newcommand{\bra}[1]{\langle #1 |}

\renewcommand{\rho}{\varrho}

\newcommand{\processnext}[1]{%
  \ifx\listfinish#1\empty\else\listact{#1}\expandafter\processnext\fi}

{\begin{framed}\begin{small}}
{\end{small}\end{framed}}

\DeclareGraphicsExtensions{.pdf,.png,.jpg}

\makeindex

\begin{document}
\title{Monogamy of Temporal Correlations: Witnessing non-Markovianity Beyond Data Processing}
\date{\today}

\author{Matheus Capela}
\affiliation{Institute of Physics, Federal University of Goi\'{a}s, POBOX 131, 74001-970, Goi\^{a}nia, Brazil}
\author{Lucas C. C\'{e}leri}
\affiliation{Institute of Physics, Federal University of Goi\'{a}s, POBOX 131, 74001-970, Goi\^{a}nia, Brazil}
\affiliation{Department of Physical Chemistry, University of the Basque Country UPV/EHU, Apartado 644, E-48080 Bilbao, Spain}
\author{Kavan Modi}
\affiliation{School of Physics \& Astronomy, Monash University, Clayton, Victoria 3800, Australia}
\author{Rafael Chaves}
\affiliation{International Institute of Physics, Federal University of Rio Grande do Norte, 59070-405 Natal, Brazil}
\affiliation{School of Science and Technology, Federal University of Rio Grande do Norte, 59078-970 Natal, Brazil}

\begin{abstract}
The modeling of natural phenomena via a Markov process --- a process for which the future is independent of the past, given the present--- is ubiquitous in many fields of science. Within this context, it is of foremost importance to develop ways to check from the available empirical data if the underlying mechanism is indeed Markovian. A paradigmatic example is given by data processing inequalities, the violation of which is an unambiguous proof of the non-Markovianity of the process. Here, our aim is twofold. First we show the existence of a monogamy-like type of constraints, beyond data processing, respected by Markov chains. Second, to show a novel connection between the quantification of causality and the violation of both data processing and monogamy inequalities. Apart from its foundational relevance in the study of stochastic processes we also consider the applicability of our results in a typical quantum information setup, showing it can be useful to witness the non-Markovianity arising in a sequence of quantum non-projective measurements.
\end{abstract}

\maketitle

\section{Introduction} 
The importance of Markov processes can hardly be overstated. In short, a process is Markovian if, in order to predict its future state, the present state contains as much information as the full previous history. That is, a Markov process keeps only information about its immediate past. Applications of it range from computer science \cite{bolch2006queueing}, causal inference \cite{hausman1999independence, janzing2016algorithmic} and statistics \cite{gilks1995markov} to social sciences \cite{lazer2009computational} and genetics \cite{franccois2006bayesian}. Within physics, random walks \cite{spitzer2013principles} and the Brownian motion \cite{chung2013lectures} are paradigmatic examples of Markov processes. In quantum information, they also play a key role \cite{fawzi2015quantum}, particularly in the understanding of open-system dynamics \cite{breuer2016colloquium}.

Mathematically, a stochastic discrete process $\left\{ X_n, t_n \in T \right\}$ is Markovian if the probability that the random variable $X_n$ takes a value $x_n$ at time $t_n \in T$, is uniquely determined, and not affected by the possible values of $X$ at previous times to $t_{n-1}$. That is,
\begin{equation}
\label{markov}
p(x_n\vert x_{n-1},\dots,x_1)=p(x_n \vert x_{n-1}), \text{ for all } t_n \in T.
\end{equation}
Given a joint probability distribution $p(x_1,\dots,x_n)$, to check if it arises from a Markovian process we have to test all the $n$ conditions expressed in \eqref{markov}. However, if the number $n$ of variables or their cardinalities (the number of possible values they can assume) is large, it is practically impossible to gather enough statistical data to reconstruct $p(x_1,\dots,x_n)$ and thus check for its Markovianity. As an example, an English text with $1000$ letters has $1000^{26}=10^{78}$ possible variations, close to the estimated number of atoms in the universe. For that reason, we often have to rely on marginal information, for instance, only the pairwise correlations between two variables. Because of this estimation limitation, we face a particular case of a marginal problem \cite{fritz2012entropic}: given some limited/marginal information, what can we conclude from the global object?

Formally, the problem is equivalent to quantifier elimination \cite{geiger1999quantifier,chaves2016polynomial}: starting from a set of constraints --in this case the positivity and normalization of probabilities plus the Markov conditions \eqref{markov}-- we aim to eliminate from our description those variables to which we do not have empirical access (e.g., correlations involving more than two variables). The same problem arises in causal inference \cite{pearl2000causality}, an extremely demanding computational task that cannot be performed beyond very few variables \cite{garcia2005algebraic}. A more treatable approach is to ask instead, what are the constraints implied by Markovianity on the entropies of the variables of interest \cite{fritz2012entropic,chaves2014inferring,weilenmann2017analysing,Budroni2016,chaves2015information}. For instance, a simple and yet fundamental result implies that Markov processes fulfil data processing inequalities, basically stating that as we move along the Markov chain the correlation between the consecutive cannot increase, but may remain constant. Data processing inequalities are a consequence of the Markov condition, imposing constraints to the correlations between the variables in different time steps, but are they the only consequence? This is the central question we explore in this article.

As we show, indeed data processing inequalities are the only consequence of the Markov condition for $n= 3$. However, for $n \geq 4$, Markovianity implies new kinds of monogamy constraints for the correlations in the Markov chain, violations of which can be seen as a device-independent test \cite{acin2007device, gallego2010device, chaves2015device} of the non-Markovianity of the underlying process. Interestingly, by employing an operational definition of divisibility \cite{simon-div}, we show that such condition is sufficient for a process satisfying all data processing and monogamy inequalities. Furthermore, we show how the violation of these new constraints can also be connected with the quantification of causal influences \cite{chaves2014inferring, janzing2013quantifying, Causal2015} among the variables. Finally, we consider a quantum information related application of this framework, showing that non-projective measurements can give rise to statistics compatible with data processing, but nonetheless can violate our new monogamy inequalities, thus witnessing the non-Markovianity arising from such quantum measurements.

The paper is organized as follows. In Sec. \ref{sec:Shannon} we review the basic toolbox to be used in this paper: the entropic approach to the marginal problem \cite{fritz2012entropic,chaves2014inferring,weilenmann2017analysing,Budroni2016}. In Sec. \ref{sec:beyond_data} we employ this framework to show that a new kind of constraint, beyond that given by data processing inequalities, arise for Markov chains with $n \geq 4$. The question regarding divisibility is addressed in Sec.~\ref{sec:divisible}. In Sec. \ref{sec:causal} we show that the violation of both the data processing and the monogamy inequalities can be given a causal interpretation. Finally, in Sec. \ref{sec:discussion} we conclude and discuss our findings. More technical details and proofs of our results can be found in the Appendix.

\section{Shannon entropies, entropy cones and marginal problems}
\label{sec:Shannon}

The Shannon entropy of a random variable $X$ is the fundamental building block in information theory \cite{yeung2008information}. It is defined as
\begin{equation*}
H(X)=-\sum_{x} p(x)\log p(x),
\end{equation*}
where the sum is taken over the support of $X$. If we are interested in the entropy of $n$ variables $X_1,\dots,X_n$, it is useful to construct the entropy vector $h$ associated with these variables as $h=\left( H(\emptyset), H(X_1),H(X_2),H(X_1,X_2),\dots,H(X_1,\dots,X_n) \right)$. That is, $h$ is a vector with $2^n$ components given by all possible entropies among $n$ variables (including the empty-set $\emptyset$ for which we define $H(\emptyset)=0$). Within this approach, a natural question is to understand which vectors in the real space $\mathbbm{R}^{2^n}$ define the entropy vectors. The first trivial constraint follows from the fact that entropies are positive quantities, that is, the entropy vector cannot have negative components. We are thus restrict to the positive-orthant of $\mathbbm{R}^{2^n}$. The second constraint comes from the realization that entropies define an unbounded convex set, the so-called entropy cone \cite{yeung2008information}. We can think of it as a hyper-dimensional and infinite ice cream cone where the tip of cone lies in the origin of our coordinate system.

Annoyingly, however, the exact structure of the real space defining entropy vectors is still not precisely known, the best general description being given by an outer approximation of the true entropy cone, the so called Shannon cone \cite{yeung2008information}. The nice thing about the Shannon cone is the fact that it is defined in terms of finitely many linear inequalities of  two types:
\begin{widetext}
\begin{eqnarray}
\label{shannon_ineq}
I(X_2:X_3 \vert X_1) \geq 0 &\rightarrow & H(X_1,X_2)+H(X_1,X_3) \geq H(X_1,X_2,X_3)+ H(X_1), \nonumber \\
H(X_1 \vert X_2,X_3) \geq 0 &\rightarrow & H(X_1,X_2,X_3) \geq H(X_2,X_3).
\end{eqnarray}
\end{widetext}
The first inequality is known as strong subadditivity (or submodularity), basically stating the positivity of the conditional mutual information. The second constraint is known as monotonicity, stating the positivity of conditional entropies or alternatively showing that the uncertainty of the whole is at least as large as the uncertainty of its parts. Given a collection of $n$ variables there are $n+\binom{n}{2}2^{n-2}$ non-redundant inequalities (that is, inequalities that cannot be obtained by combining the other inequalities) defining the Shannon cone. This minimum set of linear inequalities is known as the \textit{elemental} set of Shannon type inequalities.

Given the Shannon cone (as represented by the elemental set), we can ask the following: what are the constraints following from the elemental set on the subspace where we eliminate some the variables in the entropy vector \cite{fritz2012entropic}? As an example let us consider the following situation. Three people, seating in different rooms in such a way they cannot directly communicate (but are allowed to establish some pre-shared correlations), have to answer questions to a referee. Each of the answers of the three participants being represented by the random variables $X_1$, $X_2$ and $X_3$. However, at a given run of our experiment, a referee will only ask questions to two of them, that is, the referee does not have access to the entropy $H(X_1,X_2,X_3)$ corresponding to the event where the three participants would give their answers. This means that we have to eliminate this entropy from our description, implying that our object of interest is a new entropy vector where we trace out one of its components. Formally, the problem is equivalent to a quantifier elimination problem \cite{geiger1999quantifier}: We have a set of (linear) inequalities and we want to have the equivalent description of this set where some of the variables appearing in the inequalities have been eliminated from the problem.

Coming back to our problem. To obtain an inequality that does not depend on $H(X_1,X_2,X_3)$, we can simply sum the two Shannon type inequalities in \eqref{shannon_ineq} to obtain
\begin{equation}
\label{shannon_coins}
H(X_1,X_2)+H(X_1,X_3)\geq H(X_2,X_3)+ H(X_1),
\end{equation}
or in terms of mutual information $I(X:Y)=H(X)+H(Y)-H(X,Y)$ (a measure of correlations between $X$ and $Y$)
\begin{equation}
\label{shannon_coins_I}
I(X_1:X_2)+I(X_1:X_3)-I(X_2:X_3) \leq  H(X_1).
\end{equation}
As we can see, the simple assumption about the existence of a joint probability distribution describing the three variables already imply constraints about their pairwise correlations.

To illustrate the use of these marginal constraints,  consider dichotomic answers (yes/no questions). Also, suppose that after we run the experiment a sufficient number of times, the referee observes that all answers are unbiased ($H(X_1)=H(X_2)=H(X_3)=1$). Furthermore, answer $X_1$ is fully correlated with $X_2$ and $X_3$ (that is $I(X_1:X_2)=I(X_1:X_3)=1$); however $X_2$ and $X_3$ are uncorrelated ($I(X_2:X_3)=0$). If we plug in these values in \eqref{shannon_coins_I} we see that this inequality is violated. What does this mean? Notice that \eqref{shannon_coins} follows from the assumption that even if we cannot observe it, there is a well defined $H(X_1,X_2,X_3)$ joint entropy for all the answers. The violation of the inequality shows that this assumption is not valid: From the marginal observations we cannot construct a well-defined $H(X_1,X_2,X_3)$ (in such a way that all the elemental inequalities are respected). In fact, notice that the distribution violating the inequality is a bit odd. Since $X_2$ and $X_3$ are fully correlated with $X_1$, by the transitivity of correlations we would expect that $X_2$ and $X_3$ are also fully correlated. We can understand this distribution as a violation of the rules of the game. At each run, parties 2 and 3 communicate through a secret channel: If in a given particular run one of them is excluded of the game (the referee does not ask one of them any question), then the other part (which was asked something) will use a strategy that correlates his answer with answer $X_1$; if both parties 2 and 3 are asked questions, they just give completely uncorrelated answers between them.

\section{Markov processes beyond Data Processing inequalities}
\label{sec:beyond_data}
In the case of a Markov process we can follow a similar construction to the one delineated above. We have $n$ variables respecting the usual elemental inequalities; however, in this case, we also have a set of new constraints that follow from the Markov conditions \eqref{markov}. In terms of entropies, \eqref{markov} can be expressed as
\begin{equation}
\label{markov_ent}
H(X_n \vert X_{n-1},\dots,X_1 )= H(X_n \vert X_{n-1}) \text{, for all } n.
\end{equation}
That is, in this case we are interested in the intersection of the Shannon cone with the hyperplanes defined by \eqref{markov_ent}. We can proceed with the quantifier elimination and thus eliminate all terms but the marginal involving single and two-body terms. 

Let us start with the simplest possible Markov chain with $n=3$.
In this case, the only Markov condition is given by $H(X_3 \vert X_1,X_2)=H(X_3 \vert X_2)$. Performing the quantifier elimination step we observe that the only non-trivial inequalities characterizing the marginal Shannon cone are given by
\begin{eqnarray*}
I(X_1:X_2) \geq I(X_1:X_3) \ , \ I(X_2:X_3) \geq I(X_1:X_3),
\end{eqnarray*}
that is, we recover the usual data processing inequalities that we expect to hold in a Markov chain. By non-trivial, we mean inequalities that are not simple elemental inequalities (strong subadditivity and monotonicity). In this geometric perspective, data processing inequalities are nothing else than the facets of the Shannon cone intersected with the hyperplanes defining the entropic Markov conditions \eqref{markov_ent}, marginalized to the subspace where the joint entropy between the three variables has been eliminated.  For instance, the data processing inequality $I(X_1:X_2) \geq (X_1:X_3)$ is a direct consequence of combining the strong subadditivity $I(X_1:X_2 \vert X_3) \geq 0$ with the Markov condition $H(X_3 \vert X_2,X_1) = H(X_3 \vert X_2)$ (see Appendix for a simple proof).

We can now move to the case of a Markov chain with $n=4$. In this case, we have two Markov conditions: $H(X_3 \vert X_1,X_2)=H(X_3 \vert X_2)$ and $H(X_4 \vert X_1,X_2,X_3)=H(X_4 \vert X_3)$. Performing the quantifier elimination, we observe that the only non-trivial (and non-redundant) inequalities are given by
\begin{eqnarray}\label{eq:ineqset}
& & I_{1,3} \geq I_{1,4} \ , \ I_{1,2} \geq I_{1,3},\nonumber \\
& & I_{2,4} \geq I_{1,4} \ , \ I_{3,4} \geq I_{2,4}, \nonumber \\
& & I_{1,4} + I_{2,3} \geq I_{1,3} + I_{2,4},
\end{eqnarray}
where we have used the short-hand notation $I_{i,j} \equiv I(X_i:X_j)$. 

We highlight two aspects in the set of the inequalities above. First, we notice that there are other data processing inequalities that follow in this scenario (five more), for instance, $I_{1,2} \geq I_{1,4}$. However, all these other data processing inequalities are redundant, in the sense that they follow from the set above combined with Shannon type inequalities (see Appendix for a proof); this set is the minimum non-redundant one \footnote{Care has to be taken when searching for a process that satisfied the data processing inequalities, but not the monogamy inequality. This because some of the remaining data processing inequalities are guaranteed to hold only if the monogamy inequality holds, see App~\ref{app:ineqset} for details.}. Second, the most interesting feature comes from the fact that we have a new kind of inequality emerging for $n=4$ and that is not of the data processing type. It shows that in Markov processes not only the correlations should decrease as we move along the chain (as quantitatively expressed by the data processing inequality) but also that the pairwise correlations between the variables should respect a monogamy kind of constraint. Interestingly, as shown below there are non-Markovian processes respecting all data processing inequalities but nonetheless violate our new derived inequality, thus showing the clear relevance of it.

A natural question is whether new kind of monogamy relations will appear as we increase the size of the Markov chain. As proven in the Appendix this is indeed the case. For instance for $n=6,8,10$ the following monogamy inequalities hold
\begin{widetext}
\begin{eqnarray*}
I_{1,6} + I_{2,5} + I_{3,4} \geq  I_{1,4} + I_{2,6} + I_{3,5} \\
I_{1,8} +I_{2,7} +I_{3,6} +I_{4,5}\geq I_{1,5}+I_{2,8}+I_{3,7}+I_{4,6} \\
I_{1,10} + I_{2,9} + I_{3,8} +I_{4,7} +I_{5,6} \geq I_{1,6}+I_{2,10}+I_{3,9}+I_{4,8} + I_{5,7}.
\end{eqnarray*}    
\end{widetext}
In particular, notice that the sum of the distances between the nodes involved is the same in the left hand side (LHS) and right hand side (RHS) of the monogamy inequalities. For instance, the mutual information $I_{1,6}$ involves nodes $X_1$ and $X_6$ that have four nodes in between them. So, for $n=6$ the sum of the distances in the LHS is $4+2+0=6$, while in the RHS we have $2+3+1=6$. A similar analysis shows that for $n=8$, the sum of distances in the LHS gives $6+4+2+0=12$ and the RHS gives $3+5+3+1=12$; for $n=10$ the LHS gives $8+6+4+2+0=20$ and RHS gives $4+7+5+3+1=20$. Based on that and the clear pattern observed for $n=4,6,8,10$ we conjecture that the following monogamy inequality holds for arbitrary even $n$
\begin{equation}
\label{Mngen}
    \sum_{i=1}^{n} I_{i,n-i+1} \geq I_{1,1+n/2} + \sum_{i=2}^{n/2} I_{i,n-i+1}.
\end{equation}
This result is based on induction and an analytical general proof is missing.

\subsection{Examples of monogamy violation while satisfying the data processing inequalities}

\noindent
\textbf{Two bit process.}
We now given an explicit example of a classical non-Markovian process where the data processing inequalities hold, while the monogamy inequality does not. Consider that variable $X_1$ is binary ($x_1=0,1$) and all the others can assume four values $x_2,x_3,x_4=0,1,2,3$. The process is described by the following four steps:

\begin{enumerate}
    \item Let $x_1=0,1$ with probability $1/2$.
    
    \item If $x_1=0$, then $x_2=0,1$ with probability $1/2$.\\
    If $x_1=1$, then  $x_2=2,3$ with probability $1/2$.

    \item If $x_2=\{0,1\}$ then $x_3=0,1$ with probability $1/2$.\\
    If $x_2=\{2,3\}$ then  $x_3=2,3$ with probability $1/2$.

    \item If $(x_1,x_2,x_3)=(0,0,0) \text{ or } (1,2,2)$, then $x_4=0$. \\
    If $(x_1,x_2,x_3)=(0,1,0) \text{ or } (1,3,2)$, then $x_4=1$. \\
    If $(x_1,x_2,x_3)=(0,0,1) \text{ or } (1,2,3)$, then $x_4=2$. \\
    If $(x_1,x_2,x_3)=(0,1,1) \text{ or } (1,3,3)$, then $x_4=3$.
\end{enumerate}

This protocol generates a distribution $p(x_1,x_2,x_3,x_4)$ that with the same probability $1/8$ is equal to $(0,0,0,0)$, $(0,1,0,1)$, $(0,0,1,2)$, $(0,1,1,3)$, $(1,2,2,0)$, $(1,3,2,1)$, $(1,2,3,2)$, $(1,3,3,3)$. Computing the associated entropies we get: 
\begin{align*}
&H(X_1)=1, \ H(X_2) = H(X_3) = H(X_4)=2,\\ &H(X_1,X_2) = H(X_1,X_3)=2, \mbox{ and} \notag \\ &H(X_1,X_4) = H(X_2,X_3) = H(X_2,X_4) = H(X_3,X_4)=3.\notag
\end{align*}
We can check that this distribution respects all eight data processing inequalities (the four in \eqref{eq:ineqset} plus the four redundant ones) but nonetheless violates the monogamy constraint in \eqref{eq:ineqset}. Notice that the process is Markovian among the three first nodes ($X_2$ is a function of $X_1$ alone, and $X_3$ is a function of $X_2$ alone). However, $X_4$ has a direct dependence on the values of $X_2$ and $X_3$ (thus breaking the Markov condition).

\vspace{0.2cm}
\noindent
\textbf{One bit process.}
Consider the one bit process described by the following joint distribution
\begin{eqnarray*}
&&p(X_1,X_2,X_3,X_4)=\\ \notag
&& \quad \frac{1}{100} (3, 9, 10, 9, 6, 9, 2, 1, 1, 5, 11, 4, 4, 8, 6, 12),
\end{eqnarray*}
where the first entry in the vector corresponds to $p(0,0,0,0)$, the second to $p(0,0,0,1)$ until the last entry corresponding to $p(1,1,1,1)$. For this process all data processing inequalities are satisfied, while the monogamy is violated by $-0.0027$. This example was found by random numerical search.

\vspace{0.2cm}
\noindent
\textbf{Non-Markovianity from non-projective measurements.} We now discuss another example where the data processing inequalities are satisfied but the monogamy inequality is violated. Here the non-Markovianity arises from a sequence of non-projective quantum measurements. The generalized quantum measurements are defined by a collection of measurement operators $M_x$ satisfying the completeness relation $\sum_x M_x^{\dagger}M_x = \mathbb{1}$, where each index $x$ is associated with an experimental outcome and $\mathbb{1}$ is the identity operator on the system's Hilbert space. Here we consider the case of generalized measurements performed on a qubit system. The first collection of measurement operators, written in the computational basis, are defined by
\begin{eqnarray*}
A_1 &=& \begin{bmatrix} +0.4953 + \mathrm{i} \, 0.0687 & +0.0874 -  \mathrm{i} \, 0.2751 \\ +0.2751 +  \mathrm{i} \, 0.0874 & +0.1327 +  \mathrm{i} \, 0.2564 \end{bmatrix}, \\
A_2 &=& \begin{bmatrix}  +0.1327 +  \mathrm{i} \, 0.2564 & 0.2751 + \mathrm{i} \,0.0874 \\ +0.0874 - \mathrm{i} \,0.2751 & +0.4953 + \mathrm{i} \, 0.0687 \end{bmatrix}, \\
A_3 &=& \begin{bmatrix} +0.1327 + \mathrm{i} \, 0.2564 & -0.2751 - \mathrm{i} \, 0.0874  \\ -0.0874 + \mathrm{i} \, 0.2751 & +0.4953 + \mathrm{i} \, 0.0687 \end{bmatrix}, \\
A_4 &=& \begin{bmatrix} +0.4953 + \mathrm{i} \, 0.0687 & -0.0874 + \mathrm{i} \, 0.2751 \\ -0.2751 - \mathrm{i} \, 0.0874 & +0.1327 + \mathrm{i} \, 0.2564  \end{bmatrix}.
\end{eqnarray*}
These measurement operators were chosen such that each $A_i^{\dagger}A_i \, (i=1,2,3,4)$ has the same eigenvalues, 0.3943 and 0.1057, but the associated eigenvectors are different.

The second set of measurement operators considered here are defined by
\begin{eqnarray*}
B_1 &=& \sqrt{\frac{1+\alpha}{2}}\ket{+}\bra{+}+\sqrt{\frac{1-\alpha}{2}}\ket{-}\bra{-}, \\
B_2 &=& \sqrt{\frac{1-\alpha}{2}}\ket{+}\bra{+}+\sqrt{\frac{1+\alpha}{2}}\ket{-}\bra{-},
\end{eqnarray*}
where $\ket{\pm} = (\ket{0}\pm\ket{1})/\sqrt{2}$ and the real parameter varies as $0\leq \alpha \leq 1$. For $\alpha = 1$, the operators describe a projective measurement, while a weak one is described when $\alpha \ll 1$. 

\begin{figure}[t!]
    \centering
    \includegraphics[width=0.5\textwidth]{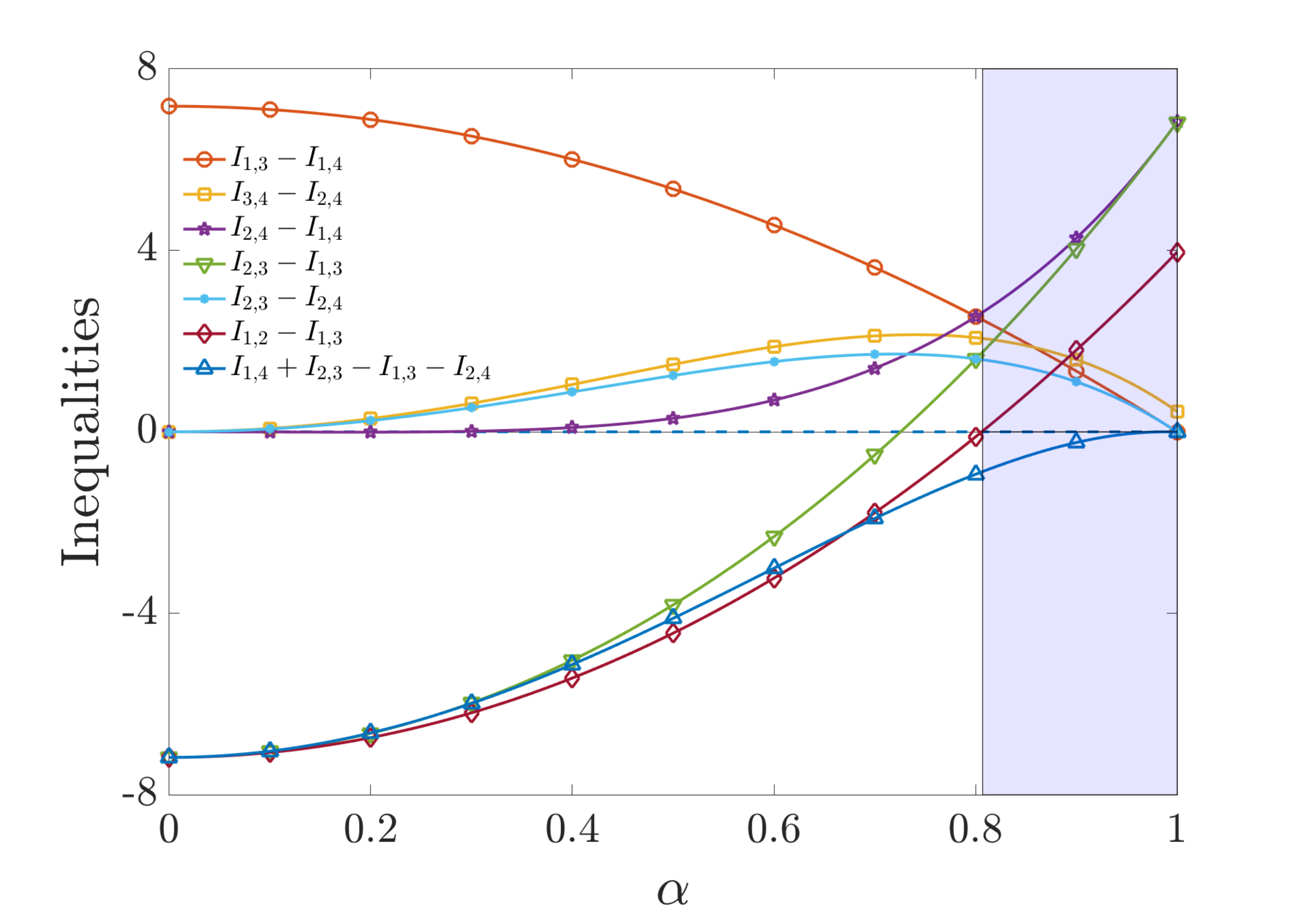}
    \caption{\textbf{Monogamy violation in series of non-projective measurements}. Data processing and monogamy inequalities discussed in the main text, given in Eqs. (\ref{eq:ineqset}), as a function of the parameter $\alpha$ (defining the measurement operator). The shaded region shows the values of $\alpha$ for which all data processing inequalities are satisfied (are positive) while the monogamy inequality is violated (is negative). There are 9 data processing inequalities for $n=4$ but only 6 of them have to be considered here (see Appendix for more details).} 
    \label{fig:monogamy}
\end{figure}

The protocol considered here is defined by four sequential measurements after the preparation of a qubit state $\rho$: The first measurement is defined by $\{A_x\}_{x=1,\cdots,4}$ and the second one by $\{B_y\}_{y=1,2}$. The third and fourth measurements are just a repetition of the first and second ones. Such sequential measurements are described by the operators $M_{i,j,k,l}=B_lA_kB_jA_i \, (i,k=1,2,3,4$ and $j,l=1,2)$, associated with the joint probability
\begin{equation}\label{eq:qprocess}
    p(i,j,k,l)= \Tr[M_{i,j,k,l} \rho M_{i,j,k,l}^{\dagger}].
\end{equation}
If the initial state is $\rho= \ket{+}\bra{+}$ it is found that the monogamy inequality is violated in the region greater than $\alpha \simeq 0.8$, but none of the data processing inequalities are violated, as shown in Fig. \ref{fig:monogamy}.

This example is interesting for several reasons. First, while the underlying process is quantum, it leads to a classical process described by the distribution in Eq. \eqref{eq:qprocess}. Secondly, the quantum process itself here is Markovian~\cite{PhysRevLett.120.040405}. We can think of it simply as the identity channel or we can think the measurements in the computational basis and the process as a unitary transformations between the measurements. In both cases the quantum process is Markovian, but it leads to a non-Markovian classical distribution. Therefore the non-Markovianity must arises from the measurements themselves. In the classical domain, making coarse measurements can turn a Markov process into a non-Markov process, see Examples 4-6 in \cite{PhysRevA.99.042108}. In this example the coarseness comes from the fact that the chosen quantum measurements are not sharp, i.e., rank-1 projections. Finally, the example illustrates a key difference between classical stochastic processes and quantum stochastic processes: In the former theory there is an assumption of non-invasiveness, while the latter requires invasive measurements to say anything about the process \cite{kolmogorov}.

\section{Inequalities for divisible processes}
\label{sec:divisible}

Divisible processes form a special superset of Markov processes. Before we talk about divisible processes we first introduce the notion of stochastic matrices. Consider the process from $X \to Y$; the stochastic matrix $\Gamma_{Y:X}$ maps any initial distribution $p(X)$ to the corresponding final distribution $p(Y)$. The stochastic matrix can be acquired from the joint probability distribution $p(X,Y)$ as
\begin{equation*}
    \Gamma_{Y:X} =
    \begin{pmatrix}
    \frac{p(X=0,Y=0)}{p(X=0)} & \frac{p(X=1,Y=0)}{p(X=1)} & \dots\\
    \frac{p(X=0,Y=1)}{p(X=0)} & 
    \frac{p(X=1,Y=1)}{p(X=1)} & \dots\\
    \vdots & 
    \vdots & \ddots
    \end{pmatrix}.
\end{equation*}

Let us now consider the process $X \to Y$ with the stochastic matrix $\Gamma_{Y:X}$ and the process $X \to Z$ with the stochastic matrix $\Gamma_{Z:X}$. A process is called divisible if matrix $G_{Z:Y} := \Gamma_{Z:X} \Gamma_{Y:X}^{-1}$ is also a stochastic matrix~\cite{NMrev}. This is known as divisibility by inversion. There is also a stronger operational notion of divisibility. Above, we have only used joint distributions $p(X,Y)$ and $p(X,Z)$. If we also have access to $p(Y,Z)$ then we can compute $\Gamma_{Z:Y}$ and check if
$\Gamma_{Z:X} = \Gamma_{Z:Y} \Gamma_{Y:X}$. The latter condition is stronger than the former because $\Gamma_{Z:X}$ represents the actual process $Y \to Z$ (see \cite{simon-div} for details). 

Importantly, it is known that there are non-Markovian processes that are divisible \cite{simon-div}. This is because divisibility only accounts for pairwise correlations and neglects higher-order correlations in time. Then a natural question is whether divisible processes satisfy the set of inequalities presented, for instance, in \eqref{eq:ineqset}? Recently, the equivalence between a non-entropic data processing inequality and divisibility was proved in Ref. \cite{PhysRevA.93.012101}. It is important to stress that their data processing inequality is distinct from the ones considered here. Furthermore, Ref. \cite{PhysRevA.93.012101} only considers divisibility by inversion, and the equivalence may not hold for when operational divisibility is considered. Here we show that the operational divisibility is sufficient for satisfying the data processing and the monogamy relations.

Suppose a process is operationally divisible; then for any map $\Gamma_{Z:X}$ can be written as $\Gamma_{Z:X} = \Gamma_{Z:Y} \Gamma_{Y:X}$ for some intermediate time step $Y$. Here, the RHS is a Markov process, meaning all pairwise correlations, i.e., the mutual information $\{I(Z:X), I(Y:X), \dots \}$ can be obtained from the underlying Markov process. Since the inequalities in Eq. \eqref{eq:ineqset} only requires pairwise correlations, which in effect come from a Markov process, all inequities there will be satisfied because they are derived under the Markov assumption. It is worth stating that the same argument does not hold if the process is divisible by inversion.

\subsection{Example relating divisibility and the inequalities}

\noindent
\textbf{A divisible non-Markovian process.} We give here an illustrative example of a non-Markovian process that is divisible and thus only carries higher-order correlations. Consider a one bit process with $x_1=0,1$ with probability $1/2$. Let $x_j = y_j$ for $j=2,3$, where $y_j$ are random bits. Finally, we let $x_4=x_1+y_2+y_3$ (here we have modular addition). It's clear that the mutual information between any two marginals will be zero since they are all independent and random. In fact, the process is divisible and therefore all inequalities in \eqref{eq:ineqset} will be trivially satisfied. However, higher-order correlations, those containing correlations between three or more variables, cannot be obtained in the same way. Indeed this is exactly why a divisible process can be non-Markovian. For instance, the mutual information such as $I(X_4 : X_1 X_2 X_3)$ or $I(X_4 : X_1 | X_2 X_3)$ will not vanish.

\vspace{0.2cm}
\noindent
\textbf{An indivisible non-Markovian process that satisfies all inequalities.} It should be now clear that there are non-Markovian processes that have correlations that cannot be detected by the inequalities in \eqref{eq:ineqset}. In fact, the converse to above statement does not hold in general. That is, there are non-divisible (hence non-Markovian) processes that satisfy all of the inequalities in \eqref{eq:ineqset}. As an example, consider the correlated process
\begin{eqnarray}
&&p(X_1,X_2,X_3,X_4) = \\ \notag
&&\quad \frac{1}{100}(6, 9, 6, 6, 2, 1, 4, 10, 1, 9, 8, 8, 4, 10, 10, 6).
\end{eqnarray}
Here we find that the monogamy inequality and all data processing inequalities hold.

\section{Causal interpretation of the data processing and monogamy inequalities}
\label{sec:causal}
The violation of the data processing or monogamy inequalities imply that the Markov the constraint is not fulfilled by the process under investigation. It seems natural that the more we violate such constraints, the more non-Markovian the process should be. In order to formalize that quantitatively, we make use here of the causal Bayesian networks formalism \cite{pearl2000causality}. 

A central concept is that of a \emph{directed acyclic graph} (DAG) which has the variables $X_i$'s as vertices. The directed edges in the DAG represent relations of causal and effect, reason why the graph should be acyclic, otherwise we would incur in paradoxical situations where the effect is its own cause. For the $X_i$'s to form a Bayesian network (with respect to the DAG), every variable should be expressed as a function of its graph-theoretical parents $PA_i$ and an unobserved noise term $N_i$ (such that the $N_i$'s are jointly independent). That is the case if and only if the distribution is of the form
\begin{equation*}
p(x) = \prod_{i=1}^n p (x_i | \mathrm{pa}_{i} ).
\end{equation*}
We notice that this is equivalent to the so-called local Markov property stating that every $X_i$ is conditionally independent of its non-descendants $ND_i$ given its parents $PA_i$, that is, $p(x_i,nd_i \vert pa_i)=p(x_i\vert pa_i)p(nd_i \vert pa_i)$.

\begin{figure}[t!]
    \centering
    \includegraphics[width=0.45\textwidth]{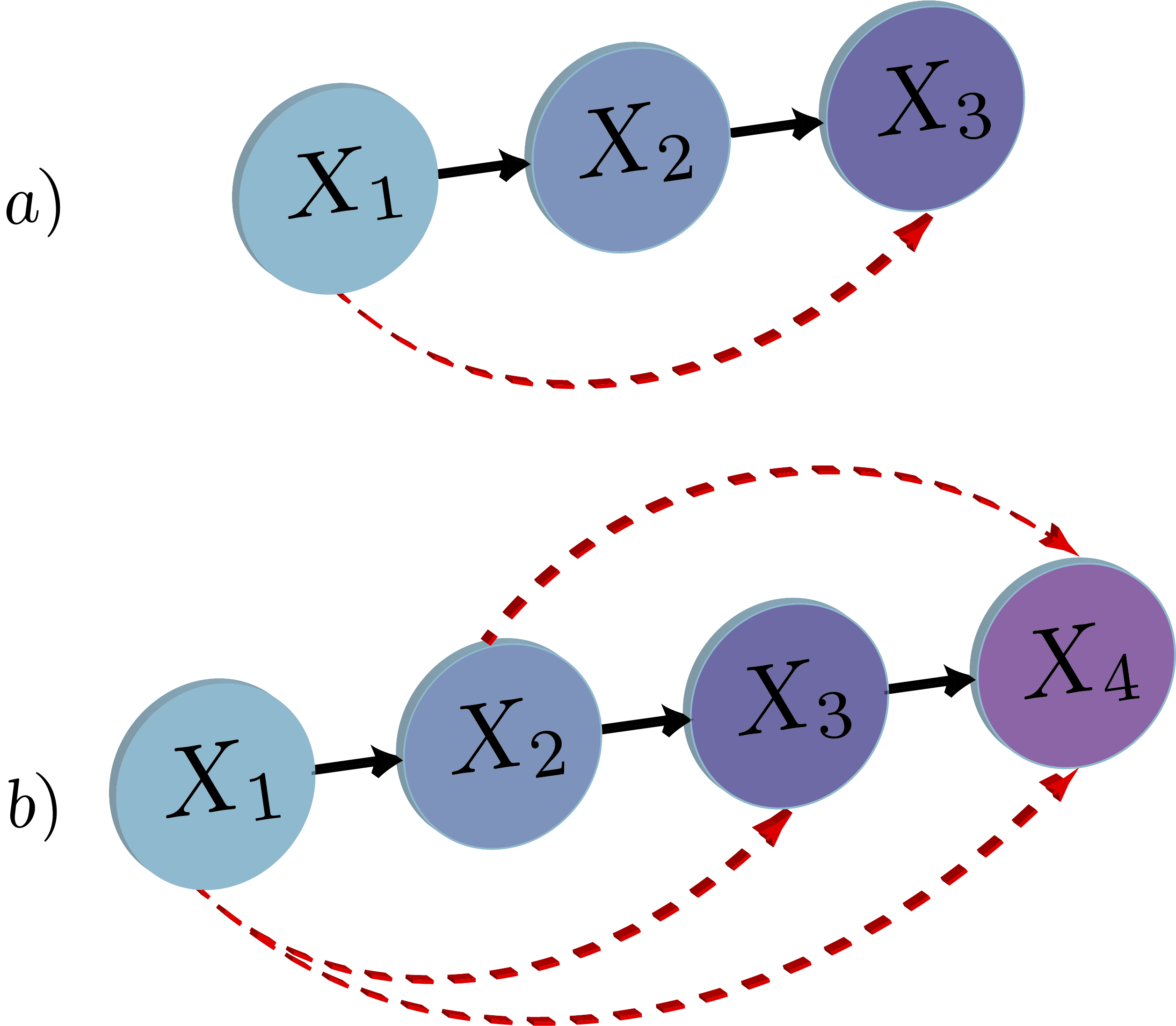}
    \caption{\textbf{DAGs illustrating a Markovian/non-Markovian chain}. A DAG with \textbf{a)} 3 and \textbf{b)} 4 nodes. The arrows in solid black represent the causal relations respected by a Markov process. In dashed red, the causal arrows imposing a violation of the Markov condition.}
    \label{fig:dags}
\end{figure}

Within this context, a Markovian process is nothing else than a causal model where a given variable $X_i$ has only $X_{i-1}$ as a parent and $X_{i+1}$ as a descendant (see Fig. \ref{fig:dags}). Thus, a natural way to quantify non-Markovianity is to quantify how much causal dependence $X_i$ has on $X_{i-2},\dots,X_{1}$. To that aim, we notice that a list of reasonable postulates any measure of causal strength should fulfill has been proposed in \cite{janzing2013quantifying}, in particular the axiom stating that
\begin{equation*}
\mathcal{C}_{X\rightarrow Y} \geq I(X : Y \vert \mathrm{PA}^X_Y)    
\end{equation*}
where $\mathcal{C}_{X\rightarrow Y}$ is the causal strength of $X$ into $Y$ and $\mathrm{PA}^X_Y$ stands for the parents of variable $Y$ other than $X$. In general, if the Markovian condition is not fulfilled, $H(X_i\vert X_{i-i},\dots,X_{1}) \neq H(X_i\vert X_{i-i})$, this means that the variables $(X_1,\dots,X_{i-2})$ have a direct causal influence over $X_i$ (see Fig. \ref{fig:dags}) and that can be quantified as $\mathcal{C}_{ (X_1,\dots,X_{i-2}) \rightarrow X_i} \geq I(X_1,\dots,X_{i-2}:X_i \vert X_{i-1}) $.

To draw a connection between causality, entropies and Markovianity, we first notice that if a process is Markovian then
\begin{equation}
\label{eq:Markovjoint}
    H(X_1,\cdots,X_n):= \sum_{i=1}^{n-1} H(X_i,X_{i+1})-\sum_{i=2}^{n-2}H(X_i),
\end{equation}
as can be seen by a direct application of the chain rule plus the Markov conditions \eqref{markov_ent}. A violation of \eqref{eq:Markovjoint} implies that at least one of the Markov conditions \eqref{markov_ent} is not fulfilled. In other terms, there is a causal influence over some $X_i$ that is not simply given by its immediate past variable $X_{i-1}$. Nicely, it can be demonstrated (see Appendix for the proof) that the following inequality holds for the sum of all non-Markovian causal influences:
\begin{widetext}
\begin{equation*}
\sum_{i=3,\dots,n}\mathcal{C}_{(X_1,\dots,X_{i-2}) \rightarrow X_{i}}  \geq \sum_{i=1}^{n-1}H(X_i,X_{i+1}) 
- \sum_{i=2}^{n-2}H(X_i) 
- H(X_1,\cdots,X_n),
\end{equation*}
\end{widetext}
where $n \geq 3$. The right hand side is equal to zero if and only if the entropic Markov constraints \eqref{markov_ent} are fulfilled. Therefore, the sum of causal influences equals to zero if and only if the entropies of the stochastic process are indeed Markovian. The more the Markov equality \eqref{eq:Markovjoint} is violated, the more non-Markovian causal influences should be present in the underlying process.

To further illustrate the connection between causality and Markovianity, consider the data processing gap, expressed as
\begin{equation}\label{eq:gap}
\mathrm{DP^1_{23}} = -I_{1,2} + I_{1,3},
\end{equation}
which is always negative for Markov processes and may be positive for non-Markovian processes. A violation of it implies that $\mathrm{DP^1_{23}} >0 $ and thus $H(X_3\vert X_2,X_1) \neq H(X_3\vert X_2)$. Using the entropic framework described above we can prove that (see Appendix)
\begin{equation*}
\mathcal{C}_{X_1 \rightarrow X_3} \geq  \mathrm{DP^1_{23}},
\end{equation*}
thus showing that the more we violate the data processing inequality the more direct causal effect the variable $X_1$ has over $X_3$.

Similarly, considering the monogamy gap for $n=4$, expressed as
\begin{equation*}
\mathrm{M_4} = - I_{1,4} - I_{2,3} + I_{1,3} + I_{2,4},
\end{equation*}
is also negative for all Markov processes and can be positive for some non-Markovian process. As proven in the Appendix, the monogamy inequality imposes a lower bound to the total non-Markovianity of the process, as quantified by the sum of causal influences
\begin{equation}
\label{causalM4}
\mathcal{C}_{(X_1,X_2) \rightarrow X_4} +\mathcal{C}_{X_1 \rightarrow X_3} \geq \mathrm{M_4}.
\end{equation}
We could analytically prove (see Appendix), up to $n=10$, that the following inequality holds
\begin{equation*}
\sum_{i=3,\dots,n}\mathcal{C}_{(X_1,\dots,X_{i-2}) \rightarrow X_{i}}  \geq M_n,
\end{equation*}
where, similarly to $M_4$ defined above, $M_n$ is the sum of mutual information terms appearing in the monogamy inequalities \eqref{Mngen}. Unfortunately, a general result for larger chains is still missing. Overall, this framework allows for an operational causal interpretation for the violation of data processing and monogamy inequalities, as a lower bound for the non-Markovian causal influences necessary to explain the observed correlations.

\section{Discussion}
\label{sec:discussion}

The main question addressed in this article is the following: if a process is Markovian, what are the additional restrictions (beyond data processing inequality) imposed on the pairwise correlations between the variables in the Markov chain? This is a very important question since we do not have, in general, access to the full probability distribution describing the process. For instance, we cannot perform sequential time measurements in a continuous way, implying that a sequence of events in time must necessarily be discrete. Our contribution in the search for answering this question, based on an entropic approach, is threefold.

First, we know that Markov processes satisfy all data processing inequalities, which are restrictions on the possible correlations between distinct time steps. However, the converse to this statement is not true. Along these lines, we demonstrate that data processing constraints are not the only consequences of the Markov condition. We proved a new class of relations, named monogamy inequalities, that can be violated by non-Markovian processes that satisfy all data processing inequalities. In this way, the violation of one of these monogamy inequalities can be employed as a device-independent test of the non-Markovianity of the underlying process. It is worthy to highlight that increasing the size of the Markov chain it is likely that new kinds of constraints, beyond data processing and monogamy, might appear. Also, in the derivation of the monogamy inequalities, only strong subadditivity constraints have been employed; meaning that they also hold for von Neumman entropies, if the corresponding joint density matrix respecting the Markov condition can be defined \cite{PhysRevLett.120.040405, PhysRevA.97.012127, 1367-2630-18-6-063032}. Exploring these possibilities defines a clear venue for future research.

Our second contribution deals with the concepts of Markovianity and divisibility, sometimes seem as synonymous in the literature. By considering divisible processes, we showed that the definition of operational divisibility is sufficient to guarantee that the process will satisfy all data processing and monogamy inequalities, even if the process is non-Markovian (but operationally divisible). This result is very interesting because it points out the distinction between Markovianity and divisibility, in such a way that it can be experimentally investigated.

Finally, our third contribution is to build a connection between the violation of data processing and monogamy inequalities with causal influences. In short, the more these inequalities are violated, greater will be the causal influence from the past. This implies that these relations can be employed as a quantifier for causal influences, since they satisfy all the requirements for a bona fide measure.

In summary, the ideas put forward in this article may have potential applications in several fields, like causal inference, statistical physics and information theory.

\begin{acknowledgments}
\textbf{Acknowledgements}. MC, LCC and RC thanks the funding agency CNPq (Grants No. 307172/2017-1, No. 406574/2018-9 and NO. 305740/2016-4 and INCT-IQ). RC also acknowledges the Brazilian ministries MEC and MCTIC, the Serrapilheira Institute (Grant No. Serra1708-15763) and the John Templeton Foundation via Grant Q-CAUSAL No. 61084. LCC thanks FAPEG (PRONEX \#201710267000503), CAPES (Finance Code 001) and support from Spanish MCIU/AEI/FEDER (PGC2018-095113-B-I00), Basque Government IT986-16, the projects QMiCS (820505) and OpenSuperQ (820363) of the EU Flagship on Quantum Technologies and the EU FET Open Grant Quromorphic and the U.S. Department of Energy, Office of Science, Office of Advanced Scientific Computing Research (ASCR) quantum algorithm teams program, under field work proposal number ERKJ333. KM is supported through ARC FT160100073.
\end{acknowledgments}

\bibliography{refs}

\newpage

\begin{widetext}
\appendix

\subsection{Analytical derivation of the data processing inequality}
Let us first see how the data processing inequalities can be proven. Simply use the Markov condition $H(X_3 \vert X_1,X_2)=H(X_3 \vert X_2)$ (rewritten as $H(X_1, X_2,X_3)=H(X_1, X_2)+H(X_2, X_3)-H(X_2)$) in the elemental inequality
\begin{equation*}
H(X_1,X_3)+ H(X_2,X_3) \geq H(X_1,X_2,X_3)+ H(X_3),
\end{equation*}
to obtain
\begin{equation*}
-H(X_1,X_2)+ H(X_1,X_3)+H(X_2)-H(X_3) \geq 0,
\end{equation*}
that can be rewritten as
\begin{equation*}
I_{1,2} \geq I_{1,3},
\end{equation*}
employing the same notation used in the main text.

\subsection{Analytical derivation of the monogamy inequality for $n=4$}
Let us now prove the new inequality beyond data processing. Add the following basic inequalities
\begin{eqnarray*}
H(X_1,X_2,X_4)+H(X_1,X_3,X_4) \geq H(X_1,X_2,X_3,X_4)+H(X_1,X_4) \\
H(X_1,X_2)+H(X_2,X_4) \geq H(X_1,X_2,X_4)+H(X_2) \\
H(X_1,X_3)+H(X_3,X_4) \geq H(X_1,X_3,X_4)+H(X_3)
\end{eqnarray*}
to obtain
\begin{equation*}
H(X_1,X_2)+H(X_1,X_3)-H(X_1,X_4)+H(X_2,X_4)+H(X_3,X_4)-H(X_2)-H(X_3)-H(X_1,X_2,X_3,X_4) \geq 0.
\end{equation*}
Using the Markov conditions $H(X_3 \vert X_1,X_2)=H(X_3 \vert X_2)$ and $H(X_4 \vert X_1,X_2,X_3)=H(X_4 \vert X_3)$ we can write
\begin{equation*}
H(X_1,X_2,X_3,X_4)=H(X_1,X_2)+H(X_2,X_3)+H(X_3,X_4)-H(X_2)-H(X_3).
\end{equation*}
Substituting that in the expression above we get
\begin{equation*}
H(X_1,X_3)-H(X_1,X_4)+H(X_2,X_4)-H(X_2,X_3) \geq 0,
\end{equation*}
that can be rewritten as
\begin{equation*}
-I_{1,3} + I_{1,4} - I_{2,4} + I_{2,3} \geq 0,
\end{equation*}

\subsection{Redundant Data Processing inequalities}
\label{app:ineqset}
Data processing (DP) inequalities are of the form \begin{equation*}
        I_{i,j}\geq I_{r,s}
    \end{equation*}
such that $i\geq r$ and $j\leq s$, where we consider without loss of generality $i<j$ and $r<s$ as a matter of mathematical convention. For the case $n=4$ we have 9 DP inequalities:
\begin{eqnarray*}
    I_{1,2} \geq I_{1,3}, \\
    I_{1,2} \geq I_{1,4}, \\
    I_{1,3} \geq I_{1,4}, \\
    I_{2,3} \geq I_{1,3}, \\
    I_{2,3} \geq I_{2,4}, \\
    I_{2,3} \geq I_{1,4}, \\
    I_{3,4} \geq I_{1,4}, \\
    I_{3,4} \geq I_{2,4}, \\
    I_{2,4} \geq I_{1,4}.
    \end{eqnarray*}
The 9 DP inequalities above are implied by the following 6 ones, as can be seen by just adding two of them:
    \begin{eqnarray*}
    I_{1,2} \geq I_{1,3}, \\
    I_{1,3} \geq I_{1,4}, \\
    I_{2,3} \geq I_{1,3}, \\
    I_{2,3} \geq I_{2,4}, \\
    I_{3,4} \geq I_{2,4}, \\
    I_{2,4} \geq I_{1,4}.
    \end{eqnarray*}    
In turn, this set of 6 DP inequalities is implied by the 4 DP inequalities plus the monogamy inequality in \eqref{eq:ineqset}. For instance, the DP inequality $I_{2,3} \geq I_{1,3}$ can be obtained by summing the DP inequality $I_{2,4} \geq I_{1,4}$ with the monogamy inequality. Similarly, the DP inequality $I_{2,3} \geq I_{2,4}$ can be obtained by summing the DP inequality $I_{1,3} \geq I_{1,4}$ with the monogamy inequality. Thus, the non-redundant set of DP inequalities is given by those expressed in \eqref{eq:ineqset}. 

\subsection{Causal Influences and Entropy}

Consider a stochastic process $\{X_1,\cdots,X_n\}$. One defines the joint Markov entropy by
\begin{equation*}
    H_{Markov}(X_1,\cdots,X_n):= \sum_{i=1}^{n-1} H(X_i,X_{i+1})-\sum_{i=2}^{n-2}H(X_i),
\end{equation*}
as the joint entropy of Markovian processes has the exact form above. The sum of the non-Markovian causal influences is lower bounded as follows 
\begin{equation}
    \sum_{i=3}^{n} \mathcal{C}_{(X_1,\dots,X_{i-2}) \rightarrow X_{i}} \geq  H_{Markov}(X_1,\cdots,X_n) -  H(X_1,\cdots,X_n).
\end{equation}
The proof follows by mathematical induction on the number of random variables. Suppose that the following equality is valid for $n$,
\begin{equation} \label{entprop}
    \sum_{i=1}^{n} I(X_1,\cdots,X_{i-2}:X_{i}|X_{i-1}) = H_{Markov}(X_1,\cdots,X_n) -  H(X_1,\cdots,X_n).
\end{equation}
Then, it implies that the proposition is also valid for $n+1$ as
\begin{eqnarray*}
\sum_{i=1}^{n+1} I(X_1,\cdots,X_{i-2}:X_{i}|X_{i-1}) &=& I(X_1,\cdots,X_{n-1}:X_{n+1}|X_{n}) + \sum_{i=1}^{n}I(X_1,\cdots,X_{i-2}:X_{i}|X_{i-1}) \\
&=& H(X_1,\cdots,X_n)+H(X_n,X_{n+1}) - H(X_n)-H(X_1,\cdots,X_{n+1}) \\
&&+ H_{Markov}(X_1,\cdots,X_n) -  H(X_1,\cdots,X_n) \\
&=& H_{Markov}(X_1,\cdots,X_{n+1}) -  H(X_1,\cdots,X_{n+1}).
\end{eqnarray*}
Now, since Eq. (\ref{entprop}) is true for $n=3$,
\begin{equation*}
    I(X_1:X_3|X_2)=H_{Markov}(X_1,X_2,X_3)-H(X_1,X_2,X_3),
\end{equation*}
we conclude that the proof holds also for $n\geq 3$.

\subsection{Proof of the causal interpretation of the data processing inequality}
We add the strong subaddivity constraint
\begin{equation*}
H(X_1,X_3)+H(X_2,X_3)-H(X_1,X_2,X_3)-H(X_3) \geq 0 
\end{equation*}
With the monotonicity constraint $I_{1,2} - I_{1,3} \geq 0$ to obtain
\begin{equation*}
H(X_1,X_3)+H(X_2,X_3)-H(X_1,X_2,X_3)-H(X_3)-H(X_1)-H(X_2)+H(X_1,X_2) \geq -I(X_1:X_2) ,
\end{equation*} 
that is equal to
\begin{equation*}
H(X_1,X_2)+H(X_2,X_3)-H(X_1,X_2,X_3)-H(X_2) \geq -I(X_1:X_2)+H(X_1)+H(X_3)-H(X_1,X_3),
\end{equation*} 
thus leading to the causal bound
\begin{equation*}
I(X_1:X_3\vert X_2) \geq -I_{1,2} + I_{1,3}.
\end{equation*} 

\subsection{Proof of the causal interpretation of the monogamy inequality ($n=4$)}
The sum of the non-Markovian causal influences for the case n=4 is lower bounded by
\begin{eqnarray*}
I(X_1:X_3|X_2)+I(X_1,X_2:X_4|X_3)&=&H(X_1,X_2)+H(X_2,X_3)-H(X_1,X_2,X_3)-H(X_2) \\
                             &+&H(X_1,X_2,X_3)+H(X_3,X_4)-H(X_1,X_2,X_3,X_4)-H(X_3) \\
                             &=&H(X_1,X_2)+H(X_2,X_3)-H(X_2)-H(X_3) - H(X_1,X_2,X_3,X_4),
\end{eqnarray*}
in agreement with the previous general derivation for any number of random variables.

Adding the following Strong Subadditivity inequalities

\begin{eqnarray*}
I(X_2:X_3|X_1,X_4)=H(X_1,X_2,X_4)+H(X_1,X_3,X_4)-H(X_1,X_2,X_3,X_4)-H(X_1,X_4) \geq 0, \\
I(X_1:X_4|X_2)=H(X_1,X_2)+H(X_2,X_4)-H(X_1,X_2,X_4)-H(X_2) \geq 0, \\
I(X_1:X_4|X_3)=H(X_1,X_3)+H(X_3,X_4)-H(X_1,X_3,X_4)-H(X_3) \geq 0, \\
\end{eqnarray*}
it follows the valid inequality for an arbitrary process,
\begin{eqnarray}
H(X_1,X_2)+H(X_2,X_3)+H(X_3,X_4)-H(X_2)-H(X_3)-H(X_1,X_2,X_3,X_4) \\
+H(X_2,X_4)+H(X_1,X_3)-H(X_2,X_3)-H(X_1,X_4) \geq 0.
\end{eqnarray}
Writing it in term of mutual information measures we have
\begin{equation*}
    I(X_1:X_3|X_2)+I(X_1,X_2:X_4|X_3) \geq - I_{1,4} - I_{2,3} +I_{1,3}+I_{2,4}  ,
\end{equation*}
and the demonstration is completed.

\subsection{Proof of the causal interpretation of the monogamy inequality ($n=6$)}

Adding the following Strong Subadditivity inequalities 

\begin{eqnarray*}
I(X_3:X_4|X_1,X_2,X_5,X_6) \geq 0 \\
I(X_1:X_3 | X_2,X_5,X_6) \geq 0 \\
I(X_4:X_6 | X_1,X_2,X_5) \geq 0 \\
I(X_3:X_6 | X_2,X_5) \geq 0 \\
I(X_2:X_4 | X_1,X_5) \geq 0 \\
I(X_2:X_5 | X_1,X_6) \geq 0 \\
I(X_1:X_6 | X_5) \geq 0 \\
I(X_1:X_5 | X_4) \geq 0 \\
I(X_2:X_5 | X_3) \geq 0 \\
I(X_1:X_6 | X_2) \geq 0 
\end{eqnarray*}
it is possible to demonstrate the causal inequality 
\begin{equation*}
    \sum_{i=3,\dots,6}\mathcal{C}_{(X_1,\dots,X_{i-2}) \rightarrow X_{i}}  \geq I_{1,4} + I_{2,6} + I_{3,5} - I_{1,6} - I_{2,5} - I_{3,4}.
\end{equation*}    

\subsection{Proof of the causal interpretation of the monogamy inequality ($n=8$)}

Adding the following Strong Subadditivity inequalities 
\begin{eqnarray*}
I(X_4:X_5|X_1,X_2,X_3,X_6,X_7,X_8) \geq 0 \\
I(X_1:X_4 | X_2,X_3,X_6,X_7,X_8) \geq 0 \\
I(X_5:X_8 | X_1,X_2,X_3,X_6,X_7) \geq 0 \\
I(X_4:X_8 | X_2,X_3,X_6,X_7) \geq 0 \\
I(X_3:X_8 | X_1,X_2,X_6,X_7) \geq 0 \\
I(X_1:X_3 | X_2,X_5,X_6,X_7) \geq 0 \\
I(X_2:X_5 | X_1,X_6,X_7) \geq 0 \\
I(X_3:X_5 | X_2,X_6,X_7)  \geq 0 \\
I(X_2:X_4 | X_3,X_6,X_7) \geq 0 \\
I(X_2:X_7 | X_1,X_6,X_8) \geq 0 \\
I(X_1:X_7 | X_5,X_6) \geq 0 \\
I(X_3:X_6 | X_2,X_7) \geq 0 \\
I(X_4:X_7 | X_3,X_6) \geq 0 \\
I(X_2:X_6 | X_1,X_8) \geq 0 \\
I(X_1:X_8 | X_6,X_7) \geq 0 \\
I(X_1:X_8 | X_2) \geq 0 \\
I(X_2:X_7 | X_3) \geq 0 \\
I(X_3:X_6 | X_4) \geq 0 \\
I(X_1:X_6 | X_5) \geq 0 \\
I(X_5:X_7 | X_6) \geq 0 \\
I(X_6:X_8 | X_7) \geq 0
\end{eqnarray*}
it is possible to demonstrate the causal inequality 
\begin{equation*}
\sum_{i=3,\dots,8}\mathcal{C}_{(X_1,\dots,X_{i-2}) \rightarrow X_{i}}  \geq I_{1,5} + I_{2,8} + I_{3,7} + I_{4,6} - I_{1,8} - I_{2,7} - I_{3,6} - I_{4,5}.
\end{equation*}   

\subsection{Proof of the causal interpretation of the monogamy inequality ($n=10$)}

Adding the following Strong Subadditivity inequalities 
\begin{eqnarray*}
I(X_5:X_6|X_1,X_2,X_3,X_4,X_7,X_8,X_9,X_{10}) \geq 0 \\ I(X_1;X_5|X_2,X_3,X_4,X_7,X_8,X_9,X_{10}) \geq 0 \\ I(X_6:X_{10}|X_1,X_2,X_3,X_4,X_7,X_8,X_9) \geq 0 \\ I(X_5;X_{10}|X_2,X_3,X_4,X_7,X_8,X_9) \geq 0 \\ I(X_1:X_4|X_2,X_3,X_6,X_7,X_8,X_9) \geq 0 \\ I(X_4;X_{10}|X_1,X_2,X_3,X_7,X_8,X_9) \geq 0 \\ I(X_2;X_5|X_3,X_4,X_7,X_8,X_9) \geq 0 \\ I(X_2:X_6|X_1,X_3,X_7,X_8,X_9) \geq 0 \\ I(X_4;X_6|X_2,X_3,X_7,X_8,X_9) \geq 0 \\ I(X_2;X_7|X_1,X_3,X_8,X_9,X_{10}) \geq 0 \\ I(X_5:X_9|X_3,X_4,X_7,X_8) \geq 0 \\ I(X_3;X_6|X_1,X_7,X_8,X_9) \geq 0 \\ I(X_2;X_4|X_3,X_7,X_8,X_9) \geq 0 \\ I(X_2:X_8|X_1,X_3,X_9,X_{10}) \geq 0 \\ I(X_3;X_8|X_1,X_7,X_9,X_{10}) \geq 0 \\ I(X_3;X_5|X_4,X_7,X_8) \geq 0 \\
I(X_6:X_8|X_1,X_7,X_9) \geq 0 \\ 
I(X_4:X_7|X_3,X_8,X_9) \geq 0 \\ I(X_1;X_3|X_2,X_9,X_{10}) \geq 0 \\ I(X_3;X_7|X_1,X_9,X_{10}) \geq 0 \\ I(X_1;X_8|X_7,X_9,X_{10}) \geq 0 \\
I(X_5:X_8|X_4,X_7) \geq 0 \\
I(X_6:X_9|X_1,X_7) \geq 0 \\
I(X_4:X_9|X_3,X_8) \geq 0 \\
I(X_2:X_9|X_1,X_{10}) \geq 0 \\ 
I(X_3:X_10|X_2,X_9) \geq 0 \\
I(X_1:X_9|X_7,X_{10}) \geq 0 \\
I(X_7:X_9|X_8,X_{10}) \geq 0 \\
I(X_4:X_7|X_5) \geq 0 \\ 
I(X_1:X_7|X_6) \geq 0 \\ 
I(X_3:X_8|X_4) \geq 0 \\
I(X_1:X_{10}|X_2) \geq 0 \\ 
I(X_2:X_9|X_3) \geq 0 \\
I(X_1:X_{10}|X_7) \geq 0 \\
I(X_7:X_{10}|X_8) \geq 0 \\
I(X_8:X_{10}|X_9) \geq 0 
\end{eqnarray*}
it is possible to demonstrate the causal inequality 
\begin{equation*}
\sum_{i=3,\dots,10}\mathcal{C}_{(x_1,\dots,x_{i-2}) \rightarrow x_{i}}  \geq I_{1,6} + I_{2,10} + I_{3,9} + I_{4,8} + I_{5,7} - I_{1,10} - I_{2,9} - I_{3,8} - I_{4,7} - I_{5,6}.
\end{equation*}  

\end{widetext}

\end{document}